\documentclass[aps,prl,reprint,superscriptaddress,groupedaddress,letterpaper,twocolumn]{revtex4-1}
\usepackage[english]{babel}
\usepackage{ucs}
\usepackage[utf8x]{inputenc}
\usepackage{graphicx}
\usepackage{dcolumn}
\usepackage{bm}
\usepackage{bbm}
\usepackage{amsmath}
\usepackage{amssymb}
\usepackage{mathrsfs}
\usepackage{bbold}
\usepackage{empheq}
\usepackage{amsfonts}

\usepackage{color,colordvi}
\definecolor{blue}{rgb}{0,0,1}
\definecolor{grey}{rgb}{0.6,0.6,0.6}

\newcommand{\bra}[1]{\langle #1|}
\newcommand{\ket}[1]{|#1\rangle}
\newcommand{\braket}[2]{\langle #1|#2\rangle}
\newcommand{\ketbra}[2]{\ket{#1}\!\bra{#2}}
\newcommand{\mm}[1]{\mathrm{#1}}
\newcommand{\ui}{\mathrm{i}}
\newcommand{\uf}{\mathrm{f}}
\newcommand{\abs}[1]{\left|#1\right|}

\def \hH{\hat{H}}
\def \hHad{\hat{H}_{\rm ad}}
\def \hU{\hat{U}}
\def \hW{\hat{W}}
\def \hV{\hat{V}}

\def \hHtweak{\hat{H}_{\rm c}}

\def \tphi{\tilde{\varphi}}

\begin{document} 

\title{Speeding up adiabatic quantum state transfer by using dressed states}

\author{Alexandre Baksic}
\affiliation{Department of Physics, McGill University, 3600 rue University, Montréal, Quebec H3A 2T8, Canada}
\author{Hugo Ribeiro}
\affiliation{Department of Physics, McGill University, 3600 rue University, Montréal, Quebec H3A 2T8, Canada}
\author{Aashish A. Clerk}
\affiliation{Department of Physics, McGill University, 3600 rue University, Montréal, Quebec H3A 2T8, Canada}

\date{\today}

\begin{abstract}
We develop new pulse schemes to significantly speed up adiabatic state transfer protocols. Our
general strategy involves adding corrections to an initial control Hamiltonian which harness
non-adiabatic  transitions. These corrections define a set of dressed states that the system follows
exactly during the state transfer. We apply this approach to STIRAP protocols and show that a
suitable choice of dressed states allows one to design fast protocols that do not require additional
couplings, while simultaneously minimizing the occupancy of the ``intermediate'' level.
\end{abstract}

\maketitle

\begin{figure}[t]
	\includegraphics[width=\columnwidth]{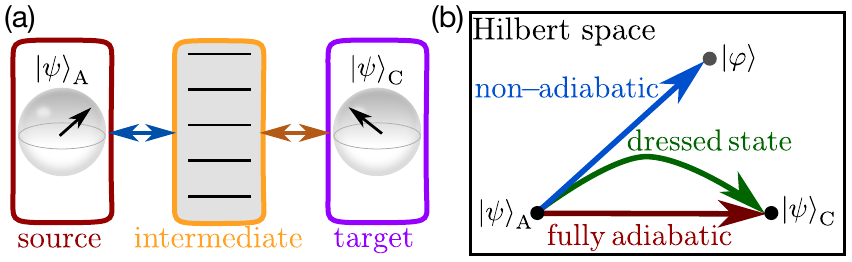}
	\caption{(Color online) (a) Schematic of a composite quantum system where the source and the
	target systems (qubits in this schematic) are coupled via some intermediate system. (b)
	Schematic of the possible evolutions : (red line) perfect adiabatic evolution, (blue line) speeding
	up the evolution results in non-adiabatic errors leading to an imperfect state transfer, (green
	line) by dressing the adiabatic eigenstates it is possible to design an evolution that leads to a
	perfect state transfer.}
	\label{fig:scheme}
\end{figure}

\textit{Introduction ---} The general goal of moving quantum states between two different systems
finds numerous applications in quantum information processing~\cite{Gisin2007,Kimble2008}. It has
generated intense theoretical interest, with numerous approaches developed to allow high
fidelity state transfer that are robust against dissipation and noise. Among the more powerful and
interesting strategies are adiabatic transfer protocols~\cite{vitanov2001}. These generically
involve adiabatically evolving an eigenstate of a composite quantum system, such that the state is
initially localized in the ``source'' system and ends up being localized in the ``target'' system
(see Fig.~\ref{fig:scheme}(a)).  The adiabatic evolution thus corresponds to a state transfer, with
the initial state of the source system ``riding'' the adiabatic eigenstates, and ending up in the
target system. The most famous examples of such approaches are the STIRAP~\cite{Bergmann1998} and
CTAP~\cite{greentree2004} protocols, well known in atomic physics. 

There are two main advantages in using transfer protocols based on adiabatic passage instead of
resonant techniques. First, adiabatic passage is inherently more robust against pulse area/timing
errors. Second, it is useful in situations where the source and target only interact via
a lossy ``intermediate'' system, as it allows one to use the mediated coupling without being harmed by the
noise. This is of particular relevance in optomechanical state transfer schemes, where a dissipative
mechanical resonator is the intermediate system~\cite{Yingdan2012,Tian2012,Dong2012,Hill2012}.  

Despite these advantages, adiabatic schemes are necessarily slow, and hence can suffer from
dissipation and noise in the target and/or source system. Therefore, several approaches have been
put forward to speed up adiabatic passage~\cite{Torrontegui2013}. Among the known methods, the
counterdiabatic control~\cite{Demirplak2003}, also referred to as transitionless
driving~\cite{Berry2009}, or its higher-order variants~\cite{Demirplak2008,Ibanez2012} are
analytical methods that allow one to construct modification of an original Hamiltonian to
compensate for non-adiabatic errors. While in principle transitionless driving would allow a perfect
state transfer, it suffers from two major flaws:~it sometimes requires either a direct coupling of
the source and target systems~\cite{Unanyan1997,chen2010,Giannelli2014,masuda2015} or a coupling not
available in the original Hamiltonian~\cite{bason2012}. The higher-order variants overcome the first
flaw of transitionless driving, but do not allow to control the population in the intermediate
system~\cite{Demirplak2008,Ibanez2012}. A related approach based on constructing dynamical
invariants~\cite{lewis1969} has also been applied to STIRAP, but it lead to pulse schemes that
either need an infinite energy gap to be perfect~\cite{Chen2012}, or do not smoothly turn on/off
\cite{Chen2012,kiely2014} and are thus extremely challenging to implement experimentally.

In this Letter, rather than constructing perfect protocols from scratch, we present an approach that
corrects existing efficient adiabatic protocols such that they allow for a perfect state transfer
even in the non-adiabatic regime. Moreover, the high flexibility of this approach allows one to
engineer and reduce the population in the intermediate lossy level. The main idea of our approach is
sketched in Fig.~\ref{fig:scheme}(b). We work with a basis of dressed states whose very definition
incorporates the non-adiabatic processes. Then, by introducing additional control fields, we can
ensure these dressed states coincide with the desired adiabatic eigenstate at the initial and final
time of the protocol. It is thus possible to do a state transfer by having the exact dynamics follow
these new dressed states, even if the protocol is too fast to allow a naive adiabatic evolution.  We
illustrate this general idea by developing simple and effective pulses for speeding up adiabatic
state transfer in generic $\Lambda$-system setups.
\\
{\it General problem ---} We consider a general composite quantum system, comprised of source,
intermediate and target subsystems, respectively labeled A, B, and C. The goal is to transfer some
initial quantum state $\ket{\psi}$ (e.g. a qubit state) from subsystem A to the target subsystem C.
Adiabatic transfer achieves this goal by constructing a time-dependent Hamiltonian whose
instantaneous eigenstates evolve in a way that facilitates the transfer. We start by assuming that
one has constructed such a protocol. The instantaneous eigenstates (hereafter referred to as
adiabatic eigenstates) and corresponding adiabatic energies (both indexed by $k$) are defined via
\begin{align}
	\hat{H}(t)\ket{\varphi_k(t)}=E_k(t)\ket{\varphi_k(t)}.
	\label{Eig_val_ad_states}
\end{align}
A subset of eigenstates has been engineered to form a basis of the A subsystem at initial time
$t_\ui$ and a basis of the target system at the final time $t_\uf$. In other words the eigenstates
$\{\ket{\varphi_{\mm{m}_j}(t)}\}_{j=0}^n$ will serve as ``medium'' states and have the following
properties:
\begin{equation}
	\ket{\varphi_{\mm{m}_j} (t_\ui)} = \ket{\beta_j}_{\mm{A}} \otimes \ket{\chi_\ui}_{\mm{B,C}},\,
	\ket{\varphi_{\mm{m}_j}(t_\uf)} = \ket{\chi_\uf}_{\mm{A,B}} \otimes \ket{\gamma_j}_{\mm{C}},	
\label{eq:finalcond}
\end{equation}
where $\{\ket{\beta_j}\}_{j=0}^n$ and $\{\ket{\gamma_j}\}_{j=0}^n$ span the subspaces A and B,
respectively. The states $\ket{\chi_\ui}_{\mm{B,C}}$ and $\ket{\chi_\uf}_{\mm{A,B}}$ are not
necessarily equal. 

It follows that if the evolution is perfectly adiabatic (i.e.~happens on a time-scale $\tau \gg
1/\Delta E$, where $\Delta E$ is the smallest instantaneous energy gap of the system), the initial
source state will be mapped on the final target state.  However for $\tau \lesssim 1/\Delta E$, the
evolution will not be perfectly adiabatic. It is convenient to move to the adiabatic frame where the
adiabatic eigenstates are time-independent.  The relevant unitary is $\hat{U}(t) =\sum_k
\ketbra{\varphi_k}{\varphi_k(t)}$.  At each instant in time, $\hat{U}(t)$ maps the adiabatic
eigenstate $\ket{\varphi_k(t)}$ onto the time-independent state $\ket{\varphi_k}$. In the adiabatic
frame, the Hamiltonian becomes:
\begin{equation}
	\hH_{\rm ad}(t) = \hH_0 (t) + \hat{W}(t) = \sum_k E_k(t)
	\ketbra{\varphi_k}{\varphi_k} + i
	\frac{d\hU(t)}{dt} \hU^{\dagger}(t)
	\label{eq:W}
\end{equation}
The operator $\hW(t)$ generically has off-diagonal matrix elements connecting the various adiabatic
eigenstates. The magnitude of these matrix elements increases as $\tau$ decreases,
leading to imperfect state transfer.
\\
\textit{Correcting non-adiabatic errors ---} In order to correct the non-adiabatic errors, we look
for a correction Hamiltonian $\hHtweak(t)$ such that the modified Hamiltonian,
$\hat{H}_{\mm{mod}} (t) = \hat{H}(t) + \hHtweak(t)$,
leads to a perfect state transfer. For this scheme to be reasonable, we require that
$\hat{H}_{\mm{mod}} (t)$ has no unattainably-large coupling strengths and that $\hHtweak(t)$ does
not involve couplings that cannot be experimentally implemented.

Our strategy is based on the observation that the corrected dynamics only needs to evolve the system
from the correct state at $t_\ui$ to the correct state at $t_\uf$ (cf. Fig.~\ref{fig:scheme}(b)).
This suggests a strategy whose crucial ingredients are:
\\ 
(I) A new basis of dressed states $\ket{\tphi_k(t)}$ formally defined by a time-dependent unitary
transformation $V(t)$ as
\begin{align}
	\ket{\tphi_k(t)} \equiv \hV (t)\ket{\varphi_k}.
\end{align}
(II) A control field $\hHtweak(t)$ that is added to the original Hamiltonian.

The additional control Hamiltonian $\hHtweak(t)$ and dressed-state basis (i.e.~$\hV(t)$) must be
chosen as to satisfy the following constraints:
\\ 
(i) The dressed medium states coincide with the medium states at time $t_\ui$ and
$t_\uf$
\begin{equation}
	\hV (t_\uf) \ket{\varphi_{\mm{m}_j}} = \hV (t_\ui) \ket{\varphi_{\mm{m}_j}} =
	\ket{\varphi_{\mm{m}_j}}.
	\label{eq:Vbdycond}
\end{equation}
(ii) For all $j$, the evolution of $\ket{\tphi_{\mm{m}_j}(t)}$ is trivial in the basis defined by $\hV (t)$.

If both these conditions are satisfied, then the perfect desired state transfer will occur. A sketch
of the general idea is shown in Fig.~\ref{fig:scheme}(b).  
\begin{figure}[t!]
	\includegraphics[width=\columnwidth]{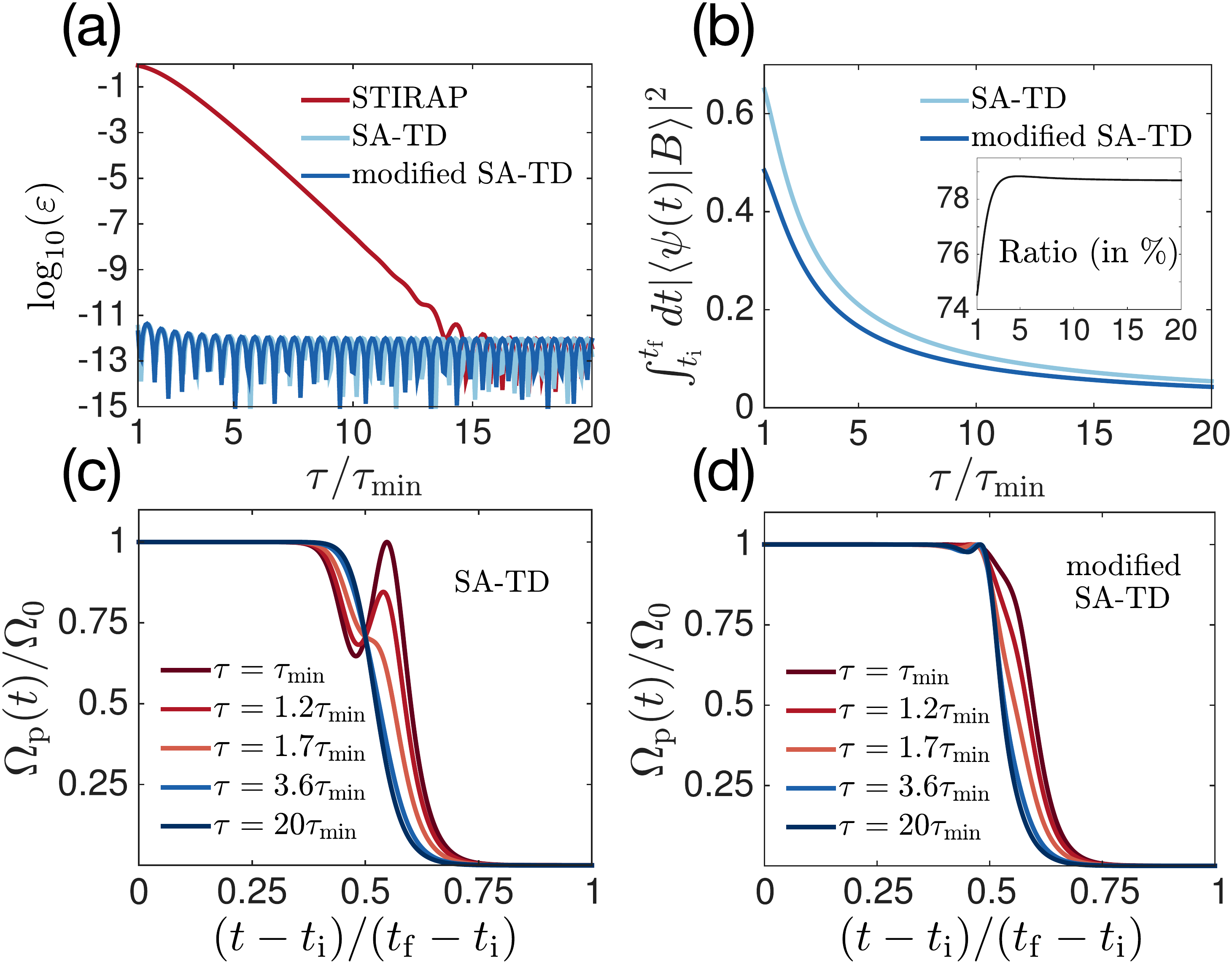}
	\caption{(Color online) (a) Comparison of the residual error between STIRAP
		Eq.~\eqref{eq:Vitanov_pulses}, SA-TD Eq.~\eqref{eq:mu_SATD}, and modified SA-TD
		Eq.~\eqref{eq:mu_gen} as a function of the effective protocol duration
		$\tau$ in units of $\tau_{\mm{min}}$. (b) Comparison of the integrated population in $\ket{B}$
		over the whole protocol time between SA-TD Eq.~\eqref{eq:mu_SATD} and our new
		dressed state approach Eq.~\eqref{eq:mu_gen} as a function of
		$\tau$ in units of $\tau_{\mm{min}}$. (Inset) Ratio of those two quantities. The integrated
		population is reduced by at least $21\%$ and at most $26\%$ with our new protocol.
		Plot of the corrected pump pulse for SA-TD (c) and modified SA-TD (d) for different
		values of $\tau$ as a function of time $(t-t_\ui)$ in units of the
		total protocol time $(t_\uf-t_\ui)$.}
\label{fig:Vitanov_protocol}
\end{figure}
Condition (ii) is better defined by moving in the frame defined by $\hV$ in which the Hamiltonian takes the form
\begin{align}
	\hH_{\rm new}(t) & = \hV \hH_{\rm ad}(t) \hV^\dagger
	+ \hV \hU \hHtweak(t) \hU^\dagger \hV^\dagger + \ i \frac{d \hV}{dt} \hV^\dagger.
	\label{eq:HSATD}
\end{align} 
We have omitted the explicit time dependence of $\hU$ and $\hV$ for clarity. Condition (ii) then
becomes
\begin{equation}
	\bra{\tphi_{\mm{m}_j}} \hH_{\rm new} \ket{\tphi_{k}} = 0 \quad
	\textrm{for } 1\leq k \leq n \,,\,  k\neq \mm{m}_j.
	\label{eq:Decoup_new}
\end{equation}
In other words, $\hHtweak(t)$ has to be designed such that it cancels the unwanted off-diagonal
elements in $\hH_{\rm new}(t)$.  

To summarize, the general method involves picking an appropriate pair of operators $(\hat{V}(t),
\hHtweak(t))$: the unitary $\hat{V}(t)$ selects a (time-dependent) basis of dressed states, while
the additional control Hamiltonian $\hHtweak(t)$ ensures the correct dynamics. The net result is
that the desired state transfer dynamics occurs perfectly despite not being in the adiabatic limit.  

Transitionless driving~\cite{Demirplak2003,Demirplak2008,Berry2009}  is a special case of this
approach and is retrieved by choosing $\hat{V}(t) = \hat{\mathbbm{1}}$ and
$\hHtweak=-\hU^{\dagger}\hW\hU$. The alternative schemes described
in~\cite{Demirplak2008,Ibanez2012} are also recovered from our approach, by choosing the dressed
states as the superadiabatic states~\cite{garrido1964,berry1987,berry1990} (instantaneous
eigenstates of $\hHad$) or its higher order counterparts. In what follows, we use our method to
derive truly new protocols.
\\
\textit{Application: STIRAP ---} We apply our general approach to the problem of Stimulated Raman
Adiabatic Passage (STIRAP)~\cite{vitanov2001,Bergmann1998} in a three-level $\Lambda$-type system.
For concreteness, each of the subsystems A, B and C are qubits such that A and C only interact with
B via the so-called pump and Stokes pulses ($\Omega_{\mm{p/s}}$ respectively).  The Hamiltonian
reads:
\begin{equation}
\begin{aligned}
	\hH(t) &= \Omega_{\mm{p}} (t) \ketbra{B}{A} + \Omega_{\mm{s}} (t) \ketbra{B}{C} + \mm{h.c.}
	\label{eq:HSTIRAP}
\end{aligned}
\end{equation}
with $\ket{A} = \ket{100}, \ket{B} = \ket{010}, \ket{C} = \ket{001}$. The pulses are parameterized
by the frequency $\Omega(t)$ and the angle $\theta(t)$
\begin{equation}
	\Omega_{\mm{p}}(t)= -\Omega(t) \sin \theta (t) \,\, , \,\, \Omega_{\mm{s}}(t)= \Omega(t) \cos \theta (t).
	\label{eq:PumpStokes}
\end{equation}
The adiabatic eigenstates (see EPAPS~\cite{EPAPS}) consist of two ``bright'' states
$\ket{\varphi_{\pm}(t)}$ with energy $E_{\pm}(t) = \pm \Omega(t)$, a ``dark'' state
$\ket{\varphi_{\mm{D}}(t)}$ with $E_{\mm{D}}(t) =0$, and $\ket{\varphi_0(t)} = \ket{000}$ with $E_0
(t) =0$. A general adiabatic state transfer from qubit A to C can be performed using the ``medium''
states
\begin{align}
	\ket{\varphi_{\mm{D}}(t)}=\cos\theta(t)\ket{A}+\sin\theta(t)\ket{C}
\end{align}
and $\ket{\varphi_0(t)}$, which operates a state transfer from $\ket{A}$ to $\ket{C}$ by using the
counter intuitive pulse sequence $\theta(t_\ui)=0$ and $\theta(t_\uf)=\pi/2$.  As mentioned before,
as the protocol time is reduced, the perfect adiabatic transfer will be more and more corrupted.
This is described by going in the adiabatic basis where the Hamiltonian \eqref{eq:HSTIRAP} becomes 
\begin{align}
	\hHad(t) =\Omega(t)\hat{M}_z+\dot{\theta}(t)\hat{M}_y,\label{eq:Had_STIRAP}
\end{align}
where $\hat{M}_z = \ketbra{\varphi_+}{\varphi_+}-\ketbra{\varphi_-}{\varphi_-}$,
$\hat{M}_x=\left(\ket{\varphi_-}-\ket{\varphi_+}\right)\!\bra{\varphi_D}/\sqrt{2}+ \mm{h.c.}$, and
$\hat{M}_y = i \left(\ket{\varphi_+}+\ket{\varphi_-}\right)\!\bra{\varphi_D}/\sqrt{2}  + \mm{h.c.}$
are spin 1 operators, obeying the commutation relation $[M_p,M_q]=i\varepsilon^{pqr}M_r$. The second
term of the adiabatic Hamiltonian Eq.~\eqref{eq:Had_STIRAP} corresponds to the non-adiabatic
couplings coming from the inertial term in Eq.~\eqref{eq:W}.

Thanks to the analogy between the adiabatic Hamiltonian \eqref{eq:Had_STIRAP} and a spin $1$ in an
magnetic field, ingredient (I) (i.e.~the construction of dressed states) of our approach can be
parametrized as a rotation of the spin with Euler angles $\xi(t)$, $\mu(t)$, and $\eta(t)$,
\begin{equation}
	\hat{V}_{\mm{g}} = \exp\left[i \eta (t) \hat{M}_z\right] \exp\left[i \mu (t) \hat{M}_x\right] 
	\exp\left[i \xi (t) \hat{M}_z\right].
	\label{eq:genU}
\end{equation}
In order to satisfy condition (i), the angle $\mu(t)$ has to satisfy $\mu(t_\ui)=\mu(t_\uf)=0(2\pi)$
and the two other angles can have arbitrary values. It can be shown that by choosing the ingredient
(II) of our method to have the general form 
\begin{align}
	\hHtweak(t)= \hat{U}^{\dagger}_{ad}(t)\left(g_x(t)\hat{M}_x+g_z(t)\hat{M}_z\right)\hat{U}_{ad}(t),
\end{align}
we find a control Hamiltonian $\hHtweak$ that does not directly couple the states $\ket{A}$ and
$\ket{C}$.  The corrected protocol will consist in a simple modification of the original STIRAP
angle and amplitude,
\begin{align}
\theta(t) & \rightarrow \tilde{\theta}(t) = \theta(t) 
	- \arctan \left( \frac{g_x(t)}{\Omega(t)+g_z(t)}\right),
	\label{eq:thetaSATD}\\
	\Omega(t) & \rightarrow	\tilde{\Omega}(t) = \sqrt{\Big(\Omega(t)+g_z(t)\Big)^2+g_x^2(t)}.
	\label{eq:OmegaSATD}
\end{align}
Moreover, in order to satisfy Eq.~\eqref{eq:Decoup_new}, the control parameters have to be chosen as
\begin{align}
&g_x(t)=\frac{\dot{\mu}}{\cos\xi}-\dot{\theta}\tan\xi,\\
&g_z(t)=-\Omega+\dot{\xi}+\frac{\dot{\mu}\sin\xi-\dot{\theta}}{\tan\mu\cos\xi},
\label{eq:control_psi}
\end{align}
and are independent of $\eta(t)$. Within our framework, it can be shown that the population in the
intermediate level $\ket{B}$ is given by
\begin{align}
	\vert\langle\psi(t)\vert B\rangle\vert^2=\sin^2\mu(t) \cos^2\xi(t).
	\label{eq:pop_B}
\end{align}
From now on, in order to keep the discussion simple, we focus on the $\xi(t)=0$ case. 
\begin{figure}[t]
	\includegraphics[width=\columnwidth]{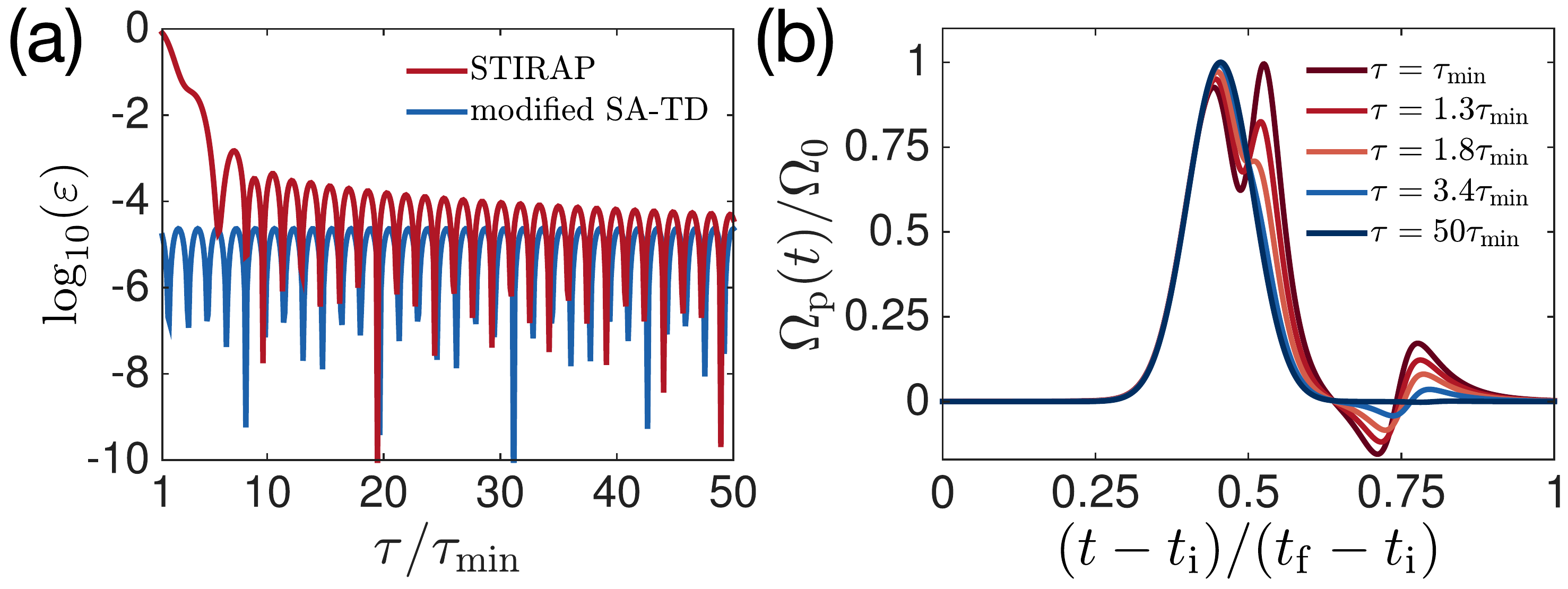}
	\caption{(Color online) (a) Comparison of the residual error for STIRAP with Gaussian densities
		Eq.~\eqref{eq:Gaussian_pulses} and modified SA-TD Eq.~\eqref{eq:mu_Gauss} as a
		function of the effective protocol duration $\tau$ in units of $\tau_{\mm{min}}$. The
		residual error is reduced by several orders of magnitude in the non-adiabatic
		regime. (b) Corrected pump pulse for different values of $\tau$ as a function of 
		time $(t-t_\ui)$ in units of the total protocol time $(t_\uf-t_\ui)$.}
	\label{fig:Gaussian_protocol}
\end{figure}
\\
\textit{Application to Vitanov-style pulses ---} We apply these dressed-state protocols to the optimal
STIRAP pulses discussed by Vitanov \emph{et al.} in Ref.~\cite{Vitanov2009} and defined by
\begin{equation}
\begin{aligned}
	\Omega(t) &  = \Omega_0 \,\, , \,\, \theta(t) & = \frac{\pi}{2} \frac{1}{1+e^{- t / \tau}},
	\label{eq:Vitanov_pulses}
\end{aligned}
\end{equation}
where the timescale $\tau$ controls the effective duration of the protocol. The simplest
nontrivial choice of dressed-states basis is the superadiabatic basis, for which 
\begin{align}
&\mu=-\arctan\left(\frac{\dot{\theta}(t)}{\Omega(t)}\right), \,\, g_x(t)=\dot{\mu}, \,\,
g_z(t)=0.
\label{eq:mu_SATD}
\end{align}
This choice will be referred to as SA-TD (superadiabatic transitionless driving). With this choice
the only way to reduce the population in the intermediate level (cf.~Eq.~\eqref{eq:pop_B}) is to
decrease the magnitude of $\dot{\theta} (t)$, and hence slow down the protocol (i.e.~longer
$\tau$). Interestingly, SA-TD represents a non-perturbative version of the DRAG approach to leakage
errors~\cite{Motzoi2009} applied to this problem (see EPAPS~\cite{EPAPS}).

Our approach allows one to construct alternatives to SA-TD (based on alternate dressed states) which
reduce the intermediate-level occupancy. By generalizing Eq.~\eqref{eq:mu_SATD} to
\begin{align}
	\mu=-\arctan\left(\frac{\dot{\theta}(t)}{f(t)\Omega(t)}\right)\!\!,\,\,g_x(t)=\dot{\mu},\,\,g_z(t)=\frac{\dot{\theta}(t)}{\tan\mu}
	\label{eq:mu_gen}
\end{align}
we can chose the auxiliary function $f(t)$ to reduce $\mu$ (and hence the amount of state dressing)
to avoid unnecessary $B$-state population.  Here, we choose to consider the simple class of
functions $f(t)=1+A\exp(-t^2/T^2)$ ($f(t)\geq 1\,\forall t$) with $A>0$ and $T>0$ two parameters
that can be optimized for each $\tau$ to minimize the population in $B$. As we show below, this
intuitive and physically motivated choice allows for a sizeable reduction of the occupancy of the
intermediate level without having to rely on more complex methods (e.g. control theory).

To compare protocols, we look at the relevant case where fidelity is limited both by a non-zero
$\tau$ in Eq.~\eqref{eq:Vitanov_pulses} and by the protocol starting and ending at a finite-time.
In theory, the protocol should start at $t_\ui=-\infty$ and end at $t_\uf=+\infty$ in order to
achieve the requirement $\theta(t_\ui)=0$, $\theta(t_\uf)=\pi/2$, and
$\mu(t_\ui)=\mu(t_\uf)=0(2\pi)$. To simulate pulses with a finite duration, we have chosen
$t_\uf=-t_\ui=15\tau$ such that $\Omega_{\mm{p}} (t_\ui) = \Omega_{\mm{s}} (t_\uf) <
10^{-6}\Omega_0$. With our choices of correction, the shorter the protocol time, the bigger the
amplitude $\tilde{\Omega}(t,\tau)$. We consider the case where each corrected pulse cannot exceed
its original maximal amplitude $\Omega_0$ ($\max_{t}
\left[\tilde{\Omega}(t,\tau)\sin\tilde{\theta}(t,\tau),
\tilde{\Omega}(t,\tau)\cos\tilde{\theta}(t,\tau)\right]\leq \Omega_0\,,\,\forall t$).  This
constraint implies that we can only correct protocols with an effective protocol time $\tau >
\tau_{\mm{min}} \simeq 1/2.63 \Omega_0$.
\begin{equation}
	\varepsilon = 1- F = 1 - \abs{_{\mm{C}}\braket{\psi (t_\uf)}{\psi (t_\ui)}_{\mm{A}}}^2.
	\label{eq:infidelity}
\end{equation}
Since we are interested in a qubit state transfer and $\ket{000}$ has a trivial dynamics, only
the transfer of state $\ket{A}$ to $\ket{C}$ gives rise to errors. Thus, we 
plot the fidelity for transferring the $\ket{A}$ state only, which sets an upper bound
for the error when transferring a superposition of an arbitrary qubit state (see EPAPS~\cite{EPAPS}). In
Fig.~\ref{fig:Vitanov_protocol}(a), we plot the residual error $\varepsilon$ as a function of $\tau$
for SA-TD Eq.~\eqref{eq:mu_SATD} and modified SA-TD Eq.~\eqref{eq:mu_gen} with optimized parameters.
Both choices reduce the residual error by the same amount and lead to several orders of magnitude
reduction as compared to the protocol defined by Eq.~\eqref{eq:Vitanov_pulses}. The oscillatory
behavior is a direct consequence of having finite-time pulses (see EPAPS~\cite{EPAPS}). 

To illustrate the additional advantage of our choice of correction, we consider the time integral over
the full protocol duration of the population in $\ket{B}$. In Fig.~\ref{fig:Vitanov_protocol}(b), we plot
this quantity for both SA-TD and modified SA-TD: the integrated population is reduced between
$\approx21-25.5\%$ with the modified SA-TD Eq.~\eqref{eq:mu_gen} as compared to SA-TD
Eq.~\eqref{eq:mu_SATD}. In Fig.~\ref{fig:Vitanov_protocol}(c) and (d), we plot the corrected pump
pulse for SA-TD and modified SA-TD for different values of $\tau$. The Stokes pulse is the symmetric
of the pump pulse with respect to $(t_\uf - t_\ui)/2$. The SA-TD pulses rapidly converge to the
Vitanov style pulses Eq.~\eqref{eq:Vitanov_pulses} when $\tau$ increases, while the modified SA-TD
pulses converge more slowly. This is due to the fact that the modified SA-TD pulses have been
designed not only to reduce the residual error, but also to reduce the population in the mechanics
which slowly converges to $0$ as $\tau\rightarrow\infty$.
\\
\textit{Application to Gaussian pulses ---} An additional advantage of our approach is that it allows
to correct protocols for which the correction Eq.~\eqref{eq:mu_SATD} does not work. In particular,
the most common approach to STIRAP uses Gaussian pulses~\cite{Bergmann1998,vitanov2001}
$\Omega_{\mm{p}} (t) = \Omega_0 \exp[-(t-t_0/2)^2/\tau^2]$ and $\Omega_{\mm{s}} (t) = \Omega_0
\exp[-(t+t_0/2)^2/\tau^2]$ with $t_0$ the delay time between the two pulses. Using the
parametrization defined in Eq.~\eqref{eq:PumpStokes}, we have
\begin{equation}
\begin{aligned}
&\theta(t)=\arctan\left[\exp(2tt_0/\tau^2)\right]\\
&\Omega(t)=\Omega_0\exp\left(-\frac{t^2+t_0^2/4}{\tau^2}\right)\sqrt{2\cosh\left(tt_0/\tau^2\right)}.
\end{aligned}
\label{eq:Gaussian_pulses}
\end{equation}
For this particular case, we cannot use the SA-TD prescription to construct a control Hamiltonian
as the condition $\mu(t_\ui)=\mu(t_\uf)=0(2\pi)$ is not satisfied (for this choice of pulse
$\dot{\theta}(t)/\Omega(t)\rightarrow +\infty$ as $t\rightarrow\pm\infty$). However, our dressed
state approach allows to find a control Hamiltonian using Eq.~\eqref{eq:control_psi} ($\xi=0$) and
\begin{equation}
	\mu(t)=-\arctan\left(\frac{\dot{\theta}(t)}{g(t)/\tau+\Omega(t)}\right).
\label{eq:mu_Gauss}
\end{equation}
Here, $g(t)/\tau$ is used to regularize $\mu(t)$: it has to be chosen such that it tends to zero at
$t_\ui$ and $t_\uf$ slower than $\dot{\theta}$.  In Fig.~\ref{fig:Gaussian_protocol}, we have
plotted the residual error for STIRAP with Gaussian densities (Eq.~\eqref{eq:Gaussian_pulses}) and
for modified SA-TD (Eq.~\eqref{eq:mu_Gauss}). We have chosen $t_0 = 6/5 \tau$ and
$g(t)=A/\cosh{\zeta t}$ with $A=1/40$ and $\zeta=9/10\tau$, which gives $\tau_{\rm
min}\approx1/1.27\Omega_0$. Under the condition $\Omega_{\mm{p}} (t_\ui) = \Omega_{\mm{s}} (t_\uf)
<10^{-6}\Omega_0$, we have $t_\uf=-t_\ui=6\tau$. This new pulse scheme leads to a reduction of the
residual error by several orders of magnitude (see Fig.~\ref{fig:Gaussian_protocol}(a)) in the
non-adiabatic regime while SA-TD Eq.~\eqref{eq:mu_SATD} fails. In Fig.~\ref{fig:Gaussian_protocol}
(b), we plot the corrected pump pulse for different values of $\tau$. The Stokes pulse is the
symmetric of the pump pulse with respect to $(\Omega_{\mm{p}},t)=(0,(t_\uf-t_\ui)/2)$.
\\
\textit{Conclusion ---} We have developed a general method to achieve a perfect state transfer
between two quantum systems coupled via an intermediate lossy system. In contrast to previous
schemes, our approach is both physically transparent and extremely flexible, allowing application to
a wide variety of realistic experimental situations.
\\
\textit{Ackowledgements ---} We acknowledge funding from the University of Chicago Quantum Engineering program and the AFOSR MURI program and H.R. acknowledges funding from the Swiss SNF.

\clearpage

\global\long\def\theequation{S\arabic{equation}}

\global\long\def\thefigure{S\arabic{figure}}

\setcounter{equation}{0}

\setcounter{figure}{0}

\setcounter{secnumdepth}{2}

\thispagestyle{empty}
\onecolumngrid

\begin{center}
{\fontsize{12}{12}\selectfont
\textbf{Supplemental \hspace{-0.2cm} Material for
	``Speeding \hspace{-0.2cm} up adiabatic quantum state transfer by using dressed states''\\ [5mm]}}
{\normalsize Alexandre Baksic, Hugo Ribeiro, and Aashish A. Clerk\\[1mm]}
{\fontsize{9}{9}\selectfont
\textit{Department of Physics, McGill University, Montréal, Quebec, Canada H3A 2T8 }}
\end{center}   
\normalsize

\section{Instantaneous eigenstates of the STIRAP Hamiltonian}
The Hamiltonian describing the interaction of the qubits A and C via the intermediate system B is given by
\begin{align}
H=\Omega(t)\Big(\cos\theta(t)\ket{C}\bra{B}-\sin\theta(t)\ket{A}\bra{B}+h.c.\Big).
\end{align}
and its instantaneous eigenstates are described by the lines of the unitary
\begin{align}
U_{\rm ad}=\left(\begin{array}{ccc} 
\sin\theta(t)/\sqrt{2} & -1/\sqrt{2} & -\cos\theta(t)/\sqrt{2}\\
\cos\theta(t) & 0 &\sin\theta(t)\\
\sin\theta(t)/\sqrt{2} & 1/\sqrt{2} & -\cos\theta(t)/\sqrt{2}
\end{array}\right),
\end{align}
where the first and last rows describe the so called ``bright'' states  ($\ket{\varphi_+}$ and
$\ket{\varphi_-}$) with instantaneous energy $E_+ (t) = +\Omega(t)$ and $E_-(t)=-\Omega(t)$,
respectively. The middle row describes the so called ``dark'' state $\ket{\varphi_{\rm D}}$ (it
does not involve the intermediate state $\ket{B}$) with a zero instantaneous energy, $E_{\mm{D}}=0$.
Since our goal is to transfer the state of qubit A to qubit C, we need to add to those eigenstate
the trivial state $\ket{000}$ which has zero energy and a trivial dynamics.
\section{Optimal $A$ and $T$ parameters for Vitanov style protocol with modified SA-TD correction}
\begin{figure}[t]
	\includegraphics[width=\columnwidth]{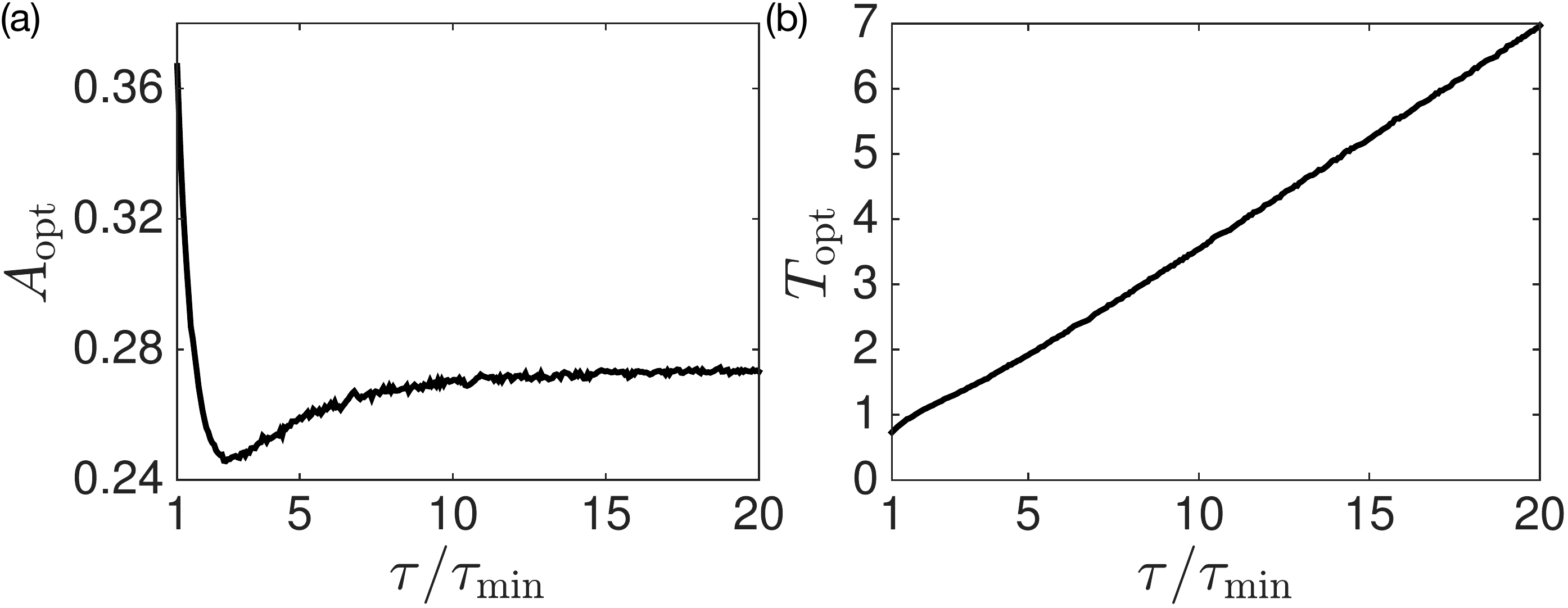}
	\caption{Optimized values of the parameters $A$ and $T$ appearing in the function $f(t)$
		Eq.~\eqref{eq:fun_A_T} as a function of $\tau/\tau_{\mm{min}}$ for $t_\ui=-t_\uf$.}
	\label{fig:Aopt_Topt}
\end{figure}

To obtain the optimal reduction of the population in the intermediate state we chose 
\begin{align}
	\mu(t)=-\arctan\left(\frac{\dot{\theta}(t)}{f(t)\Omega(t)}\right),
\end{align}
with 
\begin{align}
	f(t)=1+A\exp(-t^2/T^2)\label{eq:fun_A_T}.
\end{align}
We numerically optimize the values of $A$ and $T$ to obtain the largest reduction of the ``B'' state
population for each value of $\tau$. The results are plotted in Fig.~\eqref{fig:Aopt_Topt}.

\section{Exact dynamics of the corrected protocols}
Once the protocols are corrected, we are able to write a simple expression for the exact system
dynamics, as the correction makes the Hamiltonian diagonal in the dressed-state basis. With the
choice 
\begin{align}
&g_x(t)=\frac{\dot{\mu}}{\cos\xi}-\dot{\theta}\tan\xi,\\
&g_z(t)=-\Omega+\dot{\xi}+\frac{\dot{\mu}\sin\xi-\dot{\theta}}{\tan\mu\cos\xi},
\end{align}
we can show that 
\begin{align}
H_{\rm new}&=-\frac{\dot{\theta}+\dot{\xi}+\dot{\mu}\sin\xi}{\sin\mu\cos\xi}\hat{M}_z\\
&=\tilde{E}(t)\hat{M}_z.
\end{align}
The evolution of an arbitrary initial state is thus given by
\begin{align}
\vert\psi(t)\rangle=U^\dagger_{\rm ad}(t)V_g^\dagger(t)
\exp\left(-i\int_{t_\ui}^{t}dt'\tilde{E}(t')\hat{M}_z\right)
V_g(t_\ui)U_{\rm ad}(t_\ui)\vert\psi(t_\ui)\rangle.
\label{eq:psi_corr_ex_dyn}
\end{align}

\subsection{Fidelity and residual error of the state transfer}

\begin{figure}[t]
	\includegraphics[width=\columnwidth]{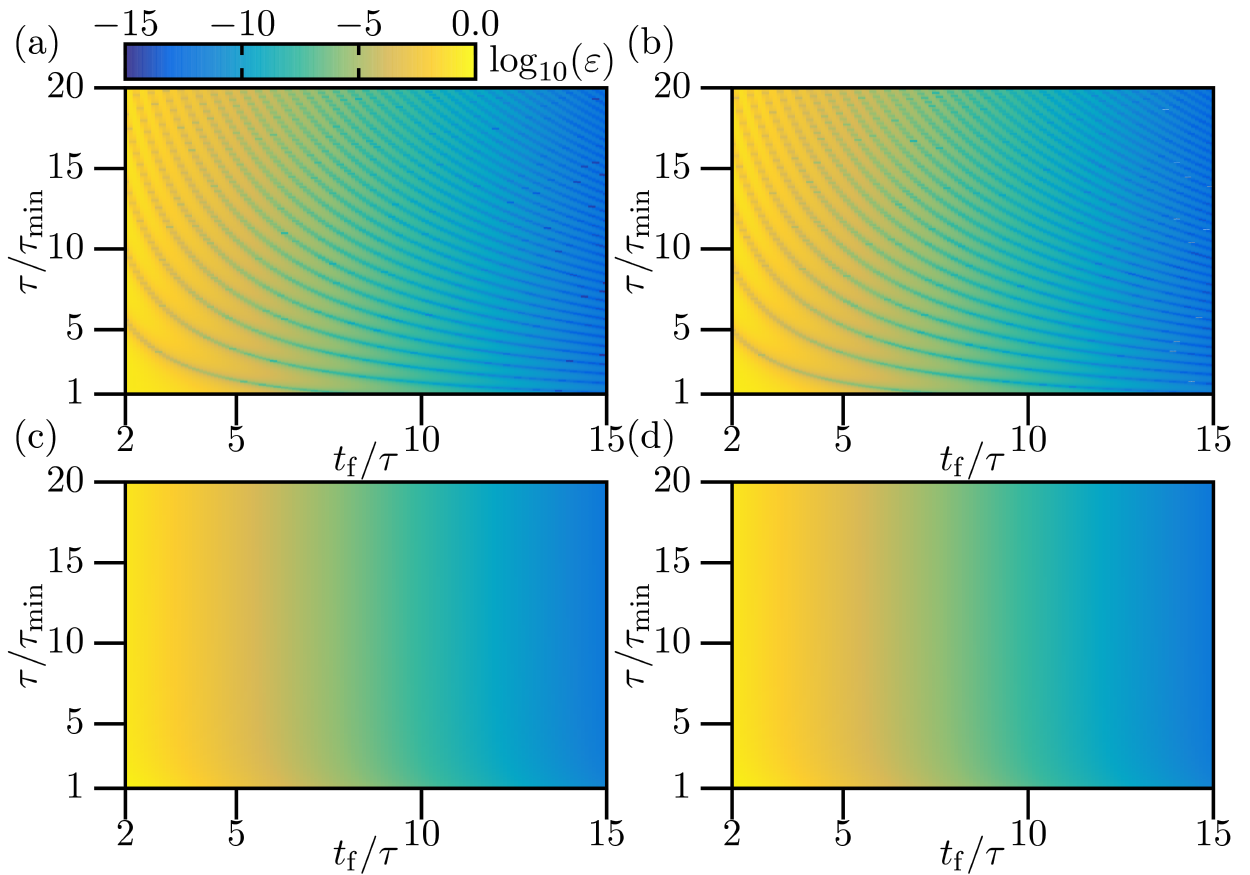}
	\caption{Error $\varepsilon=1-F$ Eq.~\eqref{eq:Fid_sym} as a function of the normalized
	effective protocol time $\tau/\tau_{\mm{min}}$ and normalized final time $t_\uf/\tau$ (we
	chose $t_\uf=-t_\ui$) for the Vitanov style corrected pulses : (a)~with SA-TD correction,
	and (b)~with modified SA-TD correction. Upper bound for the error $\varepsilon_{\rm
	max}=1-F_{\rm min}$ Eq.~\eqref{eq:Bound_F} : (c)~with SA-TD correction, and (d)~with
	modified SA-TD correction. }
	\label{fig:fid_finite_time}
\end{figure}

\begin{figure}[t]
	\includegraphics[width=\columnwidth]{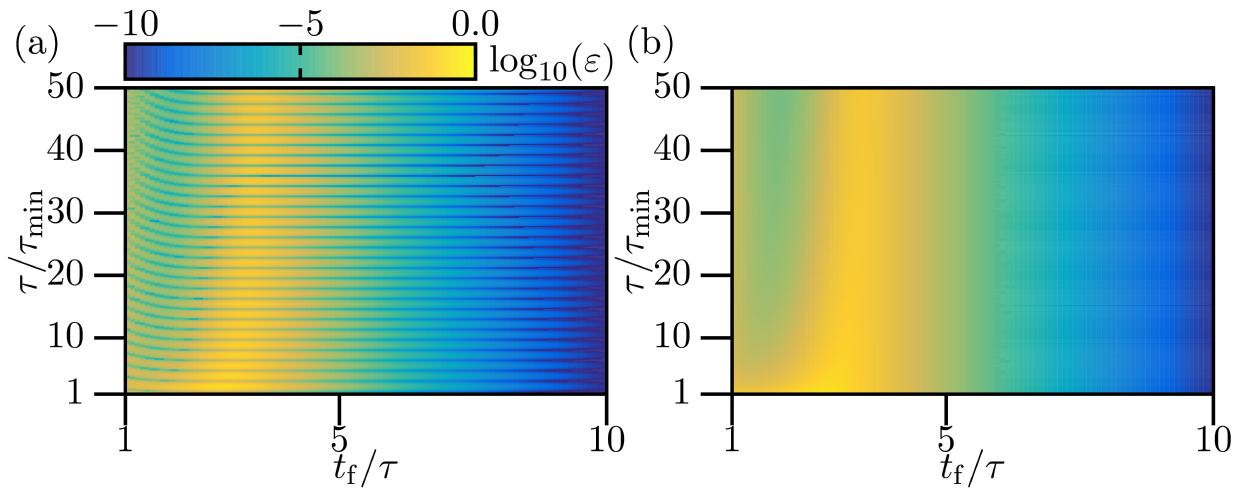}
	\caption{(a) Error $\varepsilon=1-F$ Eq.~\eqref{eq:Fid_sym} as a function of the normalized
	effective protocol time $\tau/\tau_{\mm{min}}$ and the normalized final time $t_\uf/\tau$
	(we chose $t_\uf=-t_\ui$) for the Gaussian corrected pulses. (b) Upper bound for the error
	$\varepsilon_{\rm max}=1-F_{\rm min}$ Eq.~\eqref{eq:Bound_F} for the Gaussian pulses. }
\label{fig:fid_finite_time_G}
\end{figure}

The adiabatic protocols that we have chosen (Vitanov style and Gaussian) should start at
$t_\ui=-\infty$ and end at $t_\uf=+\infty$ in order to satisfy $\theta(t_\ui)=0$ and
$\theta(t_\uf)=\pi/2$. In addition, their corrected versions (SA-TD and modified SA-TD) should
satisfy $\mu(t_\ui)=\mu(t_\uf)=0(2\pi)$ which is also done by choosing $t_\ui=-\infty$ and
$t_\uf=+\infty$. However, in a realistic setup one has a finite duration for the protocol
(both $t_\ui$ and $t_\uf$ are finite), which leads to a fidelity equal or smaller than $1$ for both
the STIRAP and corrected protocols. The fidelity of the state
transfer can be calculated with the help of Eq.~\eqref{eq:psi_corr_ex_dyn} and assuming an initial
arbitrary qubit state
$\vert\psi(t_\ui)\rangle=\cos(\alpha/2)\ket{A}+e^{i\beta}\sin(\alpha/2)\ket{000}$ . The $\ket{000}$
state has a trivial dynamics, whereas the evolution of state $\ket{A}$ is described by
Eq.~\eqref{eq:psi_corr_ex_dyn}. The fidelity is then calculated (here for $\xi(t)=0$ as in the main
text) via
\begin{align}
F=&\vert\bra{\psi(t)}C\rangle+\sin^2(\alpha/2)\vert^2\nonumber\\
=&\Big[\cos^2(\alpha/2)\Big\{
\cos\theta_\ui\sin\theta_\uf\cos\mu_\ui\cos\mu_\uf
-\cos Q(\sin\theta_\ui\cos\theta_\uf-\sin\theta_\uf\cos\theta_\ui\sin\mu_\ui\sin\mu_\uf)\nonumber\\
&+\sin Q(\sin\theta_\ui\sin\theta_\uf\sin\mu_\uf+\cos\theta_\ui\cos\theta_\uf\sin\mu_\ui)
\Big\}+\sin^2(\alpha/2)\Big]^2
\end{align}
where $\mu_{\rm k}=\mu(t_{\rm k})$, $\theta_{\rm k}=\theta(t_{\rm k})$, and
$Q=\int_{t_\ui}^{t_\uf}dt'\tilde{E}(t')$.  If we further consider symmetric protocols such that
$\theta_\uf=\pi/2-\theta_\ui$ and $\mu_\uf=\mu_\ui$ (in the Vitanov style and Gaussian pulses cases
this corresponds to $t_\ui=-t_\uf$) we obtain
\begin{align}
F=\Big[\cos^2(\alpha/2)\Big\{
\cos^2\theta_\ui\cos^2\mu_\ui
+\cos Q(\sin^2\mu_\ui\cos^2\theta_\ui-\sin^2\theta_\ui)
+\sin Q\sin\mu_\ui\sin2\theta_\ui)
\Big\}+\sin^2(\alpha/2)\Big]^2.\label{eq:Fid_sym}
\end{align}
We thus see that two oscillating terms are present ($\cos Q$ and $\sin Q$) which explains the
oscillating behaviour of the error $\varepsilon$ in the main text (see Figs. 2(a) and 3(a) in the
main text). One can calculate rigorous lower and upper bounds on
the fidelity, yielding:
\begin{equation}
F_{\rm min}(\alpha)\leq F\leq1,
\label{eq:Bound_F}
\end{equation}
with
\begin{align}
F_{\rm min}(\alpha)=\Big[\cos^2(\alpha/2)(\cos^2\theta_\ui\cos2\mu_\ui-\sin^2\theta_\ui)+\sin^2(\alpha/2)\Big]^2.
\end{align}
Since $\ket{000}$ is shared by the three subsystems (source, intermediate, and target)  and its
dynamics is trivial, it can be perfectly transferred ($\alpha=\pi(2\pi)\Rightarrow F=1$ ). The
errors due to a finite protocol time are thus due to the amplitude of $\ket{A}$ only. Hence, the
minimal fidelity is achieved for $\alpha=0(2\pi)$, which corresponds to the plots in the main text.
The error $\varepsilon$ and its maximal value $\varepsilon_{\rm max}=1-F_{\rm min}(\alpha=0)$ have
been plotted as a function of $t_\uf/\tau$ and $\tau/\tau_{\mm{min}}$ by choosing symmetric
protocols ($t_\ui=-t_\uf$) for the Vitanov style pulses in Fig.~\ref{fig:fid_finite_time} and for
the Gaussian pulses pulses in Fig.~\ref{fig:fid_finite_time_G} with $t_0=6\tau/5$.

\section{Link between DRAG-like perturbative approaches and the SA-TD approach}

It is possible to find the SA-TD correction perturbatively by using a DRAG
approach~\cite{motzoi2009,Gambetta2011}. We start from the adiabatic Hamiltonian
\begin{align}
\hat{H}_{\rm ad}=&\frac{1}{\epsilon(t)}\hat{M}_z+\dot{\theta}(t)\hat{M}_y\\
&=\frac{1}{\epsilon(t)}\hat{H}^{(-1)}+\hat{H}^{(0)}
\end{align}
where $\epsilon(t)=1/\Omega(t)$.  The second term ($\hat{H}^{(0)}$) induces leakage out of the
adiabatic subspace of interest (the subspace spanned by $\{\ket{\varphi_D},\ket{000}\}$) that we
want to eliminate at a given order in $\epsilon(t)$ by using a DRAG approach.
\\ 
The idea is to search for a unitary transformation 
\begin{align}
\hat{V}=\exp\left(-i\sum_{k=1}^{n}\epsilon^k(t)\hat{S}^{(k)}\right),
\end{align}
as well as a correction
\begin{align}
	\hat{H}_{\mm{c},\mm{ad}}=\sum_{k=1}^n \epsilon(t)^n \hat{H}^{(n)},
\end{align}
that diagonalizes the corrected Hamiltonian in the new frame
\begin{align}
	\hat{V}(\hat{H}_{\rm ad}+\hat{H}_{\mm{c},\mm{ad}})\hat{V}^\dagger+i\frac{d\hat{V}}{dt}\hat{V}^\dagger=\sum_{k=0}^\infty\epsilon^k(t)\hat{\mathcal{H}}_k
\end{align} 
at a given order $n$ in $\epsilon(t)$ (i.e. $\hat{\mathcal{H}}_k\propto \hat{M}_z\,,\,\forall k\leq
n$).  The equations that $\hat{H}^{(k)}$ and $\hat{S}^{(k)}$ have to fulfill are given in
the Appendix C of Ref.~\cite{Gambetta2011}.
\\ 
The zeroth order (in $\epsilon(t)$) equation of Appendix C in Ref.~\cite{Gambetta2011} is fullfiled by choosing 
\begin{align}
\hat{S}^{(1)}=-\dot{\theta}(t)\hat{M}_x,
\end{align}
the first order by choosing
\begin{align}
\hat{S}^{(2)}=0,\\
\hat{H}^{(1)}=-\ddot{\theta}(t)\hat{M}_x,
\end{align}
and the second order by choosing
\begin{align}
\hat{S}^{(3)}=\frac{\dot{\theta}^3(t)}{3},\\
\hat{H}^{(2)}=0.
\end{align}
The easiest way to fulfill the DRAG equations at each order is to choose $S^{(2k)}=0$, $H^{(2k)}=0$ and
$H^{(2k+1)}=d S^{(2k+1)}/dt$. By doing so at each order, one recovers the SA-TD solution :
\begin{align}
&\hat{V}(\hat{H}_{\rm ad}+\hat{H}_{c,ad})\hat{V}^\dagger+i\frac{d\hat{V}}{dt}\hat{V}^\dagger=\Omega(t)\sqrt{1+\epsilon^2(t)\dot{\theta}^2(t)}\\
&\hat{V}=\exp\left[-i\arctan\left\{\epsilon(t)\dot{\theta}\right\}\hat{M}_x\right].\\
&\hat{H}_{c,ad}=-i \hat{V}^\dagger\frac{d}{dt}\hat{V}.
\end{align}
This strategy is equivalent to first diagonalize $\hat{V}\hat{H}_{\rm ad}\hat{V}^\dagger$
perturbatively in terms of the small parameter $\dot{\theta}/\Omega(t)$ and then eliminate the
non-diagonal contributions coming from the inertial term $i\frac{d\hat{V}}{dt}\hat{V}^\dagger$ by
choosing the correction as to exactly cancel it.

\end{document}